\documentclass[conference,a4paper]{IEEEtran}
\usepackage{cite}
\usepackage[pdftex]{graphicx}
\graphicspath{./figures}
\DeclareGraphicsExtensions{.pdfsx,.jpeg,.png}
\usepackage[cmex10]{amsmath}
\usepackage{array}
\usepackage{url}
\usepackage[utf8]{inputenc}  
\usepackage[T1]{fontenc} 
\usepackage{mathtools}
\usepackage{amssymb,amsfonts}
\usepackage{dsfont}
\usepackage{dsfont}
\usepackage{stmaryrd}
\usepackage{bm}
\usepackage{citesort}

\newcommand*{\QED }{\hfill\ensuremath{\square}}

\DeclareUnicodeCharacter{00A0}{~}

\def \({\left(}
\def \){\right)}
\def \[{\left[}
\def \]{\right]}
\newcommand{\tbf}[1]{{\textbf{#1}}}

\newcommand{\defeq}{\vcentcolon=}

\newcommand{\bY}{{\textbf {Y}}}

\newcommand{\bZ}{{\textbf {Z}}}

\newcommand{\bX}{{\textbf {X}}}
\newcommand{\bx}{{\textbf {x}}}

\newcommand{\by}{{\textbf {y}}}
\newcommand{\bz}{{\textbf {z}}}

\newcommand{\bs}{{\textbf {s}}}
\newcommand{\bS}{{\textbf {S}}}

\newcommand{\be}{\begin{equation}}
\newcommand{\ee}{\end{equation}}

\newcommand\smallO{
  \mathchoice
    {{\scriptstyle\mathcal{O}}}
    {{\scriptstyle\mathcal{O}}}
    {{\scriptscriptstyle\mathcal{O}}}
    {\scalebox{.7}{$\scriptscriptstyle\mathcal{O}$}}
  }
\newcommand{\bea}{\begin{align}}
\newcommand{\eea}{\end{align}}

\newtheorem{theorem}{Theorem}[section]
\newtheorem{lemma}[theorem]{\textbf{Lemma}}
\newtheorem{thm}[theorem]{\textbf{Theorem}}
\newtheorem{remark}[theorem]{\textbf{Remark}}

\DeclareMathAlphabet{\varmathbb}{U}{bbold}{m}{n}

\newcommand{\EE}{\mathbb{E}}

\begin{document}
\title{I-MMSE relations in random linear estimation \\
and a sub-extensive interpolation method\\}
\author{\IEEEauthorblockN{Jean Barbier and Nicolas Macris}\\
\IEEEauthorblockA{Laboratoire de Théorie des Communications, Faculté Informatique et Communications,\\
Ecole Polytechnique Fédérale de Lausanne (EPFL), CH-1015 Suisse. \\
\{jean.barbier, nicolas.macris\}@epfl.ch}}
\maketitle
\IEEEpeerreviewmaketitle
\begin{abstract}
Consider random linear estimation with Gaussian measurement matrices and noise. One can compute infinitesimal variations of the mutual information under infinitesimal variations of the signal-to-noise ratio or of the measurement rate. We discuss how each variation is related to the minimum mean-square error and deduce that the two variations are directly connected through a very simple identity. The main technical ingredient is a new interpolation method called ``sub-extensive interpolation method''. We use it to provide a new proof of an I-MMSE relation recently found by Reeves and Pfister~\cite{private} when the measurement rate is varied. Our proof makes it clear that this relation is intimately related to another I-MMSE relation also recently proved in~\cite{barbier_ieee_replicaCS}. 

One can directly verify that the identity relating the two types of variation of mutual information is indeed consistent with the one
letter replica symmetric formula for the mutual information, first derived by Tanaka~\cite{Tanaka} for binary signals, and recently proved in more generality in~\cite{barbier_allerton_RLE,barbier_ieee_replicaCS,private,ReevesP16}
(by independent methods). However our proof is independent of any knowledge of Tanaka's formula. 
%
\end{abstract}
\section{Introduction}
Random linear estimation (RLE) is a fundamental research field which has been revived by a number of recent theoretical 
and practical developments such as compressed sensing~\cite{candes2006near}, error correction via sparse superposition codes~\cite{barron2010sparse}, Boolean group testing~\cite{atia2012boolean} or
code division multiple access in
communication~\cite{verdu1999spectral}. Important steps
towards a complete rigorous theory have been recently obtained. In particular the 
proof of the replica symmetric formula, a single letter formula for the asymptotic mutual 
information (MI), is now established for Gaussian RLE~\cite{barbier_allerton_RLE,barbier_ieee_replicaCS,private,ReevesP16}. 
In \cite{barbier_ieee_replicaCS} the limits of optimality of the low complexity approximate message-passing denoising algorithm are explicitly established.

An important ingredient in the 
proofs of the replica formula are interesting relations for the rate of variation (the derivative) of the MI when: $i)$
the signal-to-noise ratio varies~\cite{barbier_ieee_replicaCS}; $ii)$ the measurement rate varies~\cite{ReevesP16}. 
These formulas give the rate of variation directly in terms of the MMSE, and therefore belong to a ``family''
of I-MMSE relations, the simplest member of the family being the well known relation of Guo, Verdu and Shamai~\cite{GuoShamaiVerdu_IMMSE}. 
Of course once the replica symetric 
formula for the MI is available one can a posteriori check all these relations, however the proof of these relations does not 
involve any knowledge of the replica formula.

In this note we give a new derivation of the I-MMSE relation proved and used in~\cite{ReevesP16}. The derivation given here explicitly 
shows that all I-MMSE relations are intimately connected. The main new technical ingredient is an interpolation method, here called {\it sub-extensive interpolation method}. It involves a mix of ideas originating in interpolation methods developed in recent years for dense and sparse graphical systems, and we believe it is of 
independent interest and is bound to have applications in other problems. 

We end this introduction with a few (non-exhaustive) pointers to the literature that has led to the present work. 
The replica formula for Gaussian RLE was first proposed, on the basis of (non-rigorous) calculations using the replica method,
by Tanaka \cite{Tanaka} for the CDMA problem with
binary input signals, and was later generalized in \cite{VerduGuo2005} (see also \cite{Verdu-et-al} for recent developments in the context of compressed sensing).
Montanari and Tse~\cite{MontanariTse} sketched 
a rigorous proof of Tanaka's formula in a regime where there is no phase transition (by which we here mean no jump discontinuity in the MMSE) and 
Korada and Macris~\cite{MacrisKoradaAllerton2007,KoradaMacris_CDMA} used a Guerra-Toninelli~\cite{guerra2005introduction} interpolation method to establish that 
the replica formula is always an upper bound to the MI. In \cite{barbier_ieee_replicaCS,barbier_allerton_RLE} the converse bound (and equality)
is proven by using spatial coupling as a proof technique \cite{Giurgiu_SCproof,XXT} developped in the realm of spatially coupled 
graphical systems; namely spatial coupling for RLE~\cite{kudekar2010effect,Krzakala_CS,Andreaetalcouplingcompsensing},
threshold saturation~\cite{hassani2010coupled,kudekar2011threshold,kumar2014,pfister2012,barbier2016proof} and invariance of the
MI under spatial coupling~\cite{6284230,Giurgiu_SCproof}.

\section{Random linear estimation: Setting and results}
\label{sec:partI}
\subsection{Gaussian random linear estimation}
\label{sec:GRLE}
We consider Gaussian RLE, where one is interested in reconstructing a signal $\bs\!\in \!\mathbb{R}^N$ from 
measurements $\by\!\in \!\mathbb{R}^M$ obtained from the projection of $\bs$ by a random i.i.d Gaussian \emph{measurement matrix} $\bm{\phi}\!\in\! \mathbb{R}^{M \times N}$. 
We consider i.i.d additive white Gaussian noise (AWGN) of variance $\Delta$. Call the (standardized) noise components $Z_\mu\! \sim \!\mathcal{N}(0,1)$, $\mu\!\in\!\{1,\dots, M\}$. 
The \emph{RLE model} is
\be \label{eq:CSmodel}
\tbf{y} = \bm{\phi}\tbf{s} + \tbf{z} \sqrt{\Delta}  \ \Leftrightarrow \
y_{\mu} = \sum_{i=1}^N\phi_{\mu i} s_i + z_\mu \sqrt{\Delta}\,.
\ee

Consider a structured setting where the signal is made of $L$ i.i.d $B$-dimensional \emph{sections} $\bs_l\! \in\! \mathbb{R}^B, l\!\in\!\{1,\ldots,L\}$ distributed according to a \emph{discrete prior} $p_0(\bs_l) \defeq \sum_{k=1}^K p_k \delta(\bs_l - \tbf{a}_k)$ 
%
with a finite (but as large as desired) number of terms and all $\tbf{a}_k$'s bounded (with maximum componentwise amplitude $s_{\rm max}$). 
Thus the total number of signal components is $N\!=\!LB$. 
We denote $\bX\!\sim\!P_0$ if $\bX_l\!\sim\!p_0$ for all its i.i.d sections. 
The matrix $\bm{\phi}$ has entries $\phi_{\mu i}\!\sim\!\mathcal{N}(0,1/L)$. 
We place ourselves in the high dimensional setting where 
the \emph{measurement rate} $\alpha\!\defeq\! M/N$ is fixed when letting $L\!\to\!\infty$ ($B$ is always finite).

Define $\bar \bx \!\defeq\! \bx \!-\! \bs$, $[\bm{\phi}\bar \bx]_\mu\! \defeq \!\sum_{i=1}^N \phi_{\mu i} \bar x_i$. 
The likelihood of $\by$ is 
$P(\by|\bx)\!=\! (2\pi\Delta)^{-M/2}\exp(-\|\bm\phi\bx-\by\|^2/(2\Delta))$.
From the Bayes formula, the posterior of the RLE model is
\begin{align}
P(\bx|\by) = \frac{1}{\mathcal{Z}}\exp\Big(\!\!-\frac{1}{2\Delta}\!\sum_{\mu=1}^M \big([\bm{\phi}\bar \bx]_\mu \!-\! z_\mu \sqrt{\Delta}\big)^2 \Big)
P_0(\bx), \label{eq:posteriorCS}
\end{align}
where by a slight abuse of notation $\by$ here denotes the set of the independent \emph{quenched} random variables 
$\bm{\phi},\bs, \bz$. The normalization $\mathcal{Z}$
is the integral 
w.r.t $\bx$ of the numerator. The {\it Gibbs averages} w.r.t this posterior \eqref{eq:posteriorCS} are 
denoted by $\langle\! -\!\rangle$. For example the MMSE estimator is 
$\mathbb{E}_{\bX\vert\by}[ \bX\vert \by] \!=\! \langle \bX \rangle$. 
%
\subsection{I-MMSE relations}
\label{sec:IMMSE}
The mutual information per section $i_L \!\defeq\! i(\bY;\bX)$ is 
\begin{align} \label{eq:true_mutual_info}
i_L\!=\! \frac{1}{L}\mathbb{E}_{\bm\Phi, \bX, \bY}\Big[\!\ln\!\Big( \frac {P(\bY|\bX)}{P(\bY)}\Big)\Big]\! =\! -\frac{\alpha B}2\! 
- \!\frac{1}{L}\mathbb{E}_{\bm\Phi, \bS, \bZ}[\ln\mathcal{Z}]
\end{align}
and the MMSE per section is 
\begin{align} \label{def:mmse}
E_L \!\defeq\!  \frac{1}{L}\mathbb{E}_{\bm\Phi, \bS, \bZ}[\|\bS\! -\! \langle \bX\rangle\|^2].
\end{align}
Moreover we define a {\it measurement MMSE} as 
\begin{align}
Y_M \!\defeq \!\frac{1}{M}\mathbb{E}_{\bm\Phi, \bS, \bZ}[\|\bm{\Phi}(\bS \!-\! \langle \bX\rangle)\|^2], \nonumber
\end{align}
linked to the MI by a canonical I-MMSE relation~\cite{GuoShamaiVerdu_IMMSE,barbier_ieee_replicaCS}:
\begin{align} \label{IMMSE}
\frac{di_L}{d\Delta^{-1}}\! =\! \frac{\alpha B}{2} Y_M.
\end{align}

One can show thanks to the Guerra-Toninelli interpolation method that $i \defeq \lim_{L\to \infty} i_L$ exists (see \cite{KoradaMacris_CDMA} 
for binary signals and \cite{barbier_ieee_replicaCS} for general signals). It can also be shown from \eqref{IMMSE} that $d^2i_L/d(\Delta^{-1})^2 \leq 0$ \cite{5165186}
(indeed the measurement MMSE cannot increase by increasing the snr) so $i_L$ is a sequence of concave functions. We will repeatedly use that for a sequence of concave differentiable functions 
whose pointwise limit exists on $\mathbb{R}_+$, a standard result of real analysis states that: $i)$ the limit is concave and
continuous on all compact subsets; $ii)$ the limit is differentiable at almost every (a.e.) point; $iii)$ we can exchange the limit and derivative for a.e. points. 
Therefore for a.e. $\Delta>0$ we have $\lim_{L\to \infty} di_L/d\Delta^{-1} = di / d\Delta^{-1}$.

We now present the new I-MMSE relations specific to Gaussian RLE.
\begin{thm}[snr I-MMSE relation]\label{thm_snr_IMMSE}
For Gaussian RLE with a discrete prior, $\lim_{L\to\infty}E_L =E$ exists and for a.e. $\Delta$,
\begin{align}\label{snrMMSE}
\frac{di}{d\Delta^{-1}} = \frac{\alpha B}{2}\frac{E}{1+E/\Delta}\,.
\end{align}
\end{thm}
\begin{IEEEproof}
As remarked above 
$\lim_{L\to \infty} di_L/d\Delta^{-1}= di / d\Delta^{-1}$ for a.e. $\Delta$. Thus from \eqref{IMMSE} $(\alpha B/2)\lim_{M\to \infty} Y_M$ exists for a.e. $\Delta$ and equals $di / d\Delta^{-1}$. On the other hand 
Theorem 3.4 in \cite{barbier_ieee_replicaCS} states that a.e. $\Delta$,
\begin{align}\label{theorem34}
Y_M = \frac{E_L}{1+ E_L/\Delta} + \smallO_L(1) 
\end{align}
where $\lim_{L\to\infty}\smallO_L(1)\!=\!0$. Therefore $E\!=\!\lim_{L\to \infty}E_L$ also exists for a.e. $\Delta$ and \eqref{snrMMSE} holds. 
\end{IEEEproof}
\begin{remark}\label{premiereremarque} The proof of \eqref{theorem34} in \cite{barbier_ieee_replicaCS} (Theorem 3.4) requires concentration properties that are 
currently proven for a discrete prior. An extension to the case of a general prior (e.g., a mixture of discrete and absolutely continuous parts) could perhaps be 
obtained by quantizing the signal and showing that $\smallO_L(1)$ is uniform in the quantization but this is not immediate. A similar relation 
already appears in \cite{MontanariTse} but to the best of our knowledge the proof details are not given. 
\end{remark}

The main goal of this note is to give a new proof of the following I-MMSE relation first obtained in~\cite{private} (for $B\!=\!1$).

\begin{thm}[$\alpha$ I-MMSE relation]\label{thm_IMMSE}
For Gaussian RLE and for a.e. $\Delta$ and $\alpha$, 
\begin{align}\label{alphaMMSE}
\frac{di}{d\alpha} =\frac{B}{2} \ln\Big(1 + \frac{E}{\Delta}\Big).
\end{align}
\end{thm}
\begin{remark}\label{deuxiemeremarque} We conjecture that 
the relation is true for all $\Delta$ and a.e $\alpha$. The proof of~\cite{private} works for general priors that can have discrete and absolutely continuous parts. 
Our approach is based on concentration theorems underpinning 
\eqref{theorem34} that as explained in Remark \ref{premiereremarque} do not quite cover this general case.  
\end{remark}

We end this section by noting that eliminating $E$ from \eqref{snrMMSE} and \eqref{alphaMMSE} we obtain the interesting and simple formula
\begin{align}\label{9}
 \frac{2}{\alpha B} \frac{d i}{d\ln(\Delta^{-1})} = 1 - \exp\Big({-\frac{2}{B} \frac{d i}{d\alpha}}\Big).
\end{align}
One may check \eqref{9} directly on the replica formula for the MI.

\section{A sub-extensive interpolation method}
We introduce the sub-extensive interpolation method which allows to first prove a slightly weaker 
form of Theorem~\ref{thm_IMMSE}. All the stated lemmas are proved in the next section. 
\subsection{The interpolated perturbed model}
The interpolation is done between the RLE model \eqref{eq:CSmodel} where a measurement matrix with $M$ lines is used and one where the measurement matrix has $M\!+\!M^u$ lines, $0\!<\!u\!<\!1$. The $M^u$ additional lines have the same statistical properties and are indexed by a set ${\cal S}$ of extra indices. 
This is a sub-extensive set because $\vert \mathcal{S}\vert \!=\! M^u \!\ll\! M$.
We still denote $\by, \bm{\phi}$ the overall measurement and measurement matrix, that include the additional measurements and lines associated with ${\cal S}$.
 Define the following {\it interpolated perturbed} Hamiltonian
\begin{align} 
&\mathcal{H}_{t,h}(\bx;\by) \defeq \frac{h}{2}\sum_{i=1}^N\Big( \bar x_i  -  \frac{\widehat z_i}{\sqrt{h}}\Big)^2 +\sqrt{h}s_{\rm max}\sum_{i=1}^N |\widehat z_i| \label{pertModel}\\
&+\frac{1}{2\Delta}\sum_{\mu=1}^M \Big([\bm{\phi}\bar \bx]_\mu - z_\mu \sqrt{\Delta}\Big)^2 +\frac{t}{2\Delta}\!\sum_{\nu\in{\cal S}} \Big([\bm{\phi}\bar \bx]_\nu - z_\nu \sqrt{\frac{\Delta}{t}}\Big)^2\,.\nonumber
\end{align}
Here the interpolation parameter is $t\in[0,1]$, and going from $t\!=\!0$ to $t\!=\!1$ continuously adds the $M^u$ new measurements. 
The first perturbation term corresponds to extra measurements obtained from 
scalar AWGN ``side channels'', $y_i \!=\!s_i\!+\!\widehat z_i/\sqrt{h}$, $\widehat Z_i\!\sim\!{\cal N}(0,1)$, $i\!\in\!\{1,\ldots,N\}$, 
where the snr $h$ is ``small'' and will eventually tend to zero. This term allows to use a useful concentration result proved in 
\cite{barbier_ieee_replicaCS} (see Lemma~\ref{concentration} in sec.~\ref{proofsec} where also the second term is needed for technical reasons).

Denote the MI associated with the perturbed interpolated model $i_{t,h}$, expressed similarly to \eqref{eq:true_mutual_info} but with 
$P_{t,h}(\by|\bx)  \propto  \exp(-\mathcal{H}_{t,h}(\bx;\by))$. This leads to $i_{t,h}\!=\!  -B((1\!+\!M^{u-1})\alpha\!+\!1)/2\!-\!\mathbb{E}[\ln\mathcal{Z}_{t,h}]/L$
where $\mathcal{Z}_{t,h}\!=\!\int d\bx P_0(\bx)\exp(-\mathcal{H}_{t,h}(\bx;\by))$.
The MI of this model with and without the additional measurements is, respectively, $i_{1,h}$ and $i_{0,h}$. 
The Gibbs average $\langle\!-\!\rangle_{t,h}$ is associated with the posterior of the interpolated pertubed model 
$P_{t,h}(\bx|\by)\propto\mathcal{Z}_{t,h}^{-1}P_{t,h}(\by|\bx)P_0(\bx)$. Finally, we define 
the MMSE $E_{t,h}$ similarly as \eqref{def:mmse} but with $\langle \bX \rangle_{t,h}$ replacing 
$\langle\bX\rangle$.

\subsection{The sub-extensive interpolation}

We first show a weaker version of Theorem \ref{thm_IMMSE} for the perturbed interpolated model which is valid for all $\Delta$ but 
a.e. $h$. In sec.~\ref{finallytheproof} we show how to take the limit $h\to 0$ for a.e. $\Delta$ and thus recover Theorem \ref{thm_IMMSE}.
\begin{thm}[$\alpha$ I-MMSE relation for a.e. $h$]\label{thm_IMMSE_weak}
The following limits exist for {\it all} $\Delta$ and a.e. $h$, $\alpha$ and satisfy
\begin{align}\label{weakerequality}
\frac{d}{d\alpha}\lim_{L\to \infty}i_{0,h} = \frac{B}{2} \ln\Big(1 + \frac{\lim_{L\to \infty} E_{0,h}}{\Delta}\Big).
\end{align}
\end{thm}

The proof is based on two lemmas proved in sec.~\ref{proofsec}. 

Define a measurement MMSE associated to the subset $\mathcal{S}$:
\begin{align}\label{eq:ymmserth}
Y_{t,h}^{({\cal S})}&\defeq M^{-u} \sum_{\nu\in{\cal S}}\EE[\langle [\bm{\Phi} \bar \bX]_{\nu} \rangle_{t,h}^2 ],
\end{align}
where $\EE$ denotes the expectation w.r.t all quenched variables.
\begin{lemma}[MMSE relation]\label{lemma:MMSE}
For a.e $h$ we have
\begin{align}\label{Yresult}
Y_{t,h}^{({\cal S})}  =  \frac{E_{t,h}}{1+ E_{t,h}(t/\Delta)} + \smallO_L(1).
\end{align} 
\end{lemma}
\begin{lemma}[MMSE variation] \label{lemma:smallEdiff}
Fix $0\!<\!u\!<\!1/20$ in the interpolated perturbed model \eqref{pertModel}. Then for any $t\!\in\![0,1]$, we have $E_{t, h} = E_{0,h} + \smallO_L(1)$
for a.e. $h$.
\end{lemma}

We now sketch the proof of Theorem~\ref{thm_IMMSE_weak}.
\begin{IEEEproof}
By the fundamental theorem of calculus, one may write $i_{1,h} - i_{0,h} = \int_{0}^1 dt \,(d i_{t,h}/dt)$.
Direct differentiation gives
\begin{align}
\frac{d i_{t,h}}{dt} &= \frac{1}{2\Delta L} \sum_{\nu\in {\cal S}} \mathbb{E}\Big[\Big\langle [\bm{\Phi}\bar \bX]_{\nu}^2 - \frac{[\bm{\Phi}\bar \bX]_\nu Z_{\nu}}{\sqrt{t/\Delta}} \Big\rangle_{t,h}\Big]. 
\label{eq:A_twoterms_coup}
\end{align}
Integrating by parts over $Z_\mu\!\sim\!{\cal N}(0,1)$, \eqref{eq:A_twoterms_coup} becomes
\begin{align}
\frac{di_{t,h}}{dt} \!&=\!  \frac{\alpha BM^{u-1}}{2\Delta} Y_{t,h}^{({\cal S})} \! = \! \frac{\alpha BM^{u-1}}{2\Delta}\Big[\frac{E_{t,h}}{1\!+\! E_{t,h}(t/\Delta)} + \smallO_L(1)\Big] \nonumber
\end{align}
for a.e $h$. For the second equality we used Lemma~\ref{lemma:MMSE}. Thus
\be
i_{1,h} \!-\! i_{0,h} \! =\!  \frac{\alpha BM^{u-1}}{2\Delta} \int_{0}^{1} dt \frac{E_{t,h}}{1\!+\! E_{t,h}(t/\Delta)} +\smallO(L^{u-1}). \label{eq:diffiper_1}
\ee
This integral over $t$ cannot be calculated immediately as the MMSE depends on $t$. We overcome this difficulty using Lemma \ref{lemma:smallEdiff} and $t'\!=\!t/\Delta$. Then \eqref{eq:diffiper_1} becomes 
\begin{align}
\frac{i_{1,h} - i_{0,h}}{\alpha M^{u-1}} & =  \frac{B}{2} \int_{0}^{1/\Delta} dt' \frac{E_{0,h}}{1+ E_{0,h}t'} + \smallO_L(1)\!
 \nonumber \\
&=\frac{B}{2} \ln\Big(1 + \frac{E_{0,h}}{\Delta}\Big)+ \smallO_L(1). \label{15}
\end{align}
Note that $i_{1,h}$ is the MI of a (perturbed) RLE model with measurement rate $(M+M^u)/N=\alpha(1+\!M^{u-1})$ while $i_{0,h}$ corresponds to a measurement 
rate $\alpha$. It is then not difficult to show with
concavity inequalities w.r.t $\alpha$,\footnote{Alternatively one can also directly use the Alexandrov theorem which states that a concave function  has a second derivative almost everywhere.}
that {\it for a.e.} $\alpha$
\be
\lim_{L\to \infty}\frac{i_{1,h} - i_{0,h}}{\alpha M^{u-1}} = \lim_{L\to \infty}\frac{di_{0,h}}{d\alpha} =  \frac{d}{d\alpha}\lim_{L\to \infty}i_{0,h}.
\label{16}
\ee
Finally equations \eqref{15} and \eqref{16} imply  \eqref{weakerequality}.
\end{IEEEproof}

\subsection{Proof of Theorem \ref{thm_IMMSE}: taking the $h\to 0$ limit}\label{finallytheproof}
We consider the limit $h\to 0$ of \eqref{weakerequality}. Again, a concavity argument allows to permute this limit and the derivative for a.e. $\alpha$.
Also it is not very difficult to argue that all finite size quantities are continuous in $h\geq 0$. Therefore $\lim_{h\to 0}i_{0, h} = i_{0,0} = i_L$ and 
$\lim_{h\to 0}E_{0, h} = E_{0,0} = E_L$. So Theorem \ref{thm_IMMSE} follows if we can show that
$\lim_{h\to 0}\lim_{L\to \infty} i_{0,h} = \lim_{L\to \infty} \lim_{h\to 0} i_{0,h} = i$ and 
$\lim_{h\to 0}\lim_{L\to \infty} E_{0,h} = \lim_{L\to \infty} \lim_{h\to 0} E_{0,h} = E$ for a.e. $\Delta$. We will show that the first limit exchange is valid for all $\Delta$ and the second one for a.e. $\Delta$.

For the first limit exchange the argument is standard. 
The first derivative of $i_{0,h}$ w.r.t $h$ is an MMSE, namely
$L^{-1}\mathbb{E}[ \sum_{i=1}^N(S_i - \langle X_i\rangle_{0,h})^2]$, so its second derivative is negative because the MMSE cannot increase with increasing snr of the side channel (it can also be seen by explicit calculation \cite{barbier_ieee_replicaCS}). Thus $i_{0, h}$ is concave in $h$, and since also
$\lim_{L\to \infty} i_{0,h}$ exists, the limit is attained uniformly in $h$. This allows to exchange the limits for {\it all}
$\Delta$.

The second limit exchange is less immediate because we cannot use a convexity argument directly on the sequence 
$E_{0,h}$. However by a mild generalisation of Theorem \ref{thm_snr_IMMSE} (that follows from Lemmas 4.5 and 4.6 in \cite{{barbier_ieee_replicaCS}}) we have for a.e. $\Delta$,
\begin{align}
\frac{d}{d\Delta^{-1}}\lim_{L\to \infty} i_{0, h} 
= \frac{\alpha B}{2}\frac{\lim_{L\to \infty} E_{0, h}}{1 + \lim_{L\to \infty} E_{0, h}/\Delta}\,. \nonumber
\end{align}
Then, since $\lim_{L\to \infty} i_{0, h}$ is a concave function of $\Delta$ and its limit $h\to 0$ exists we can take the limit $h\to 0$ of this equation
and permute it with the derivative for a.e. $\Delta$. 
Thus $\lim_{h\to 0}\lim_{L\to \infty} E_{0, h}$ must exist for a.e. $\Delta$ and satisfies
\begin{align}
\frac{d i}{d\Delta^{-1}}
= \frac{\alpha B}{2}\frac{\lim_{h\to 0}\lim_{L\to \infty} E_{0, h}}{1 + \lim_{h\to 0}\lim_{L\to \infty} E_{0, h}/\Delta}\,. \nonumber
\end{align}
But since we have \eqref{snrMMSE}, we conclude $\lim_{h\to 0}\lim_{L\to \infty} E_{0, h} =E$ for a.e. $\Delta$.
Thus the limits are exchangeable for a.e. $\Delta$.
\section{Proofs of Lemmas \ref{lemma:MMSE} and \ref{lemma:smallEdiff}}\label{proofsec}

\subsection{Preliminaries}\label{prelim}
Let $\bX$, $\bX'$ two i.i.d \emph{replicas} drawn according to the product distribution $P_{t,h}(\bx|\by) P_{t,h}(\bx^\prime|\by)$. Then for any function $g$,
\begin{align}\label{nishibasic}
\EE[\langle g(\bX, \bS) \rangle_{t,h}] \!=\! \EE[\langle g(\bX, \bX')\rangle_{t,h}].
\end{align}
This identity, which has been called a Nishimori identity in the statistical mechanical literature,
follows from a simple application of Bayes formula. It has a certain number of useful consequences that we list here
(all the derivations can be found in appendix B of \cite{barbier_ieee_replicaCS}).

\subsubsection{Identity 1} First we have 
\begin{align}\label{consequence1}
2\EE[\langle[\bm{\Phi}\bar\bX]_\mu\rangle_{t,h}^2]=\EE[\langle[\bm{\Phi}\bar\bX]_\mu^2\rangle_{t,h}].
\end{align}
To derive this recall $\bar\bX = \bX-\bS$, expand the squares and systematically apply \eqref{nishibasic}. 

\subsubsection{Identity 2} Set ${\cal E}\!\defeq\! L^{-1}\sum_{i=1}^N\bar X_i X_i$. Then \eqref{nishibasic} implies 
\begin{align}\label{consequence2}
\mathbb{E}[\langle {\cal E}\rangle_{t,h}] = E_{t,h}.
\end{align}

\subsubsection{Identity 3} This one is more complicated. Define $u_{\nu} \!\defeq\! \sqrt{t/\Delta} [\bm{\phi} \bar \bx]_{\nu} \!-\! z_{\nu}$. 
From Gaussian integration by parts over $z_\nu$ and \eqref{nishibasic} one can show
\begin{align}
\EE[Z_{\nu} \langle U_\nu & \bar X_i \bar X_i'\rangle_{t,h}]
\nonumber\\
=~\!\! &\EE[ Z_{\nu}^2 S_i\langle \bar X_i\rangle_{t,h} ]\! -\! \sqrt{\frac{t}{\Delta}}\EE[ Z_{\nu}S_i\langle [\bm{\Phi}\bar \bX]_\nu \bar X_i\rangle_{t,h}].
\label{consequence3}
\end{align}

We also need the following concentration result.
\begin{lemma}[Concentration of $\cal E$]\label{concentration}
Let $\delta{\cal E}\!\defeq\!{\cal E}\!-\!E_{t,h}$ (recall \eqref{consequence2}). For any $0<a<\epsilon$ we have  
\begin{align}\label{concentr}
\int_{a}^{\epsilon} dh\, \EE[\langle \delta {\cal E}^2\rangle_{t,h}] = {\cal O}(L^{-1/10}).
\end{align}
\end{lemma}

The proof of this lemma 
is the same as the one of Proposition 8.1 in \cite{barbier_ieee_replicaCS}. This type of result is also found in 
\cite{KoradaMacris_CDMA} for binary signals. Lebesgue's dominated convergence theorem applied to 
 \eqref{concentr} implies
$\EE[\langle \delta {\cal E}^2\rangle_{t,h}] \!=\! \smallO_L(1)$ for a.e. $h\!>\!0$. 

\subsection{Proof of Lemma~\ref{lemma:MMSE}}
%

Using \eqref{eq:ymmserth} and an integration by parts w.r.t $z_\nu$ gives
\be
Y_{t,h}^{({\cal S})}=M^{-u}\sum_{\nu\in \mathcal{S}} \mathbb{E}\Big[\Big\langle [\bm{\Phi}\bar \bX]_{\nu}^2- \sqrt{\frac{\Delta}{t}}[\bm{\Phi}\bar \bX]_{\nu} Z_{\nu} \Big\rangle_{t,h}\Big], \nonumber
\ee
which combined with \eqref{consequence1} leads to
\begin{align}
&Y_{t,h}^{({\cal S})}= M^{-u}\sqrt{\frac{\Delta}{t}}\sum_{\nu\in {\cal S}}\EE[Z_{\nu}\langle [\bm{\Phi}\bar \bX]_{\nu}\rangle_{t,h}]. \label{eq:ident_ymmse}
\end{align}
Integrating by part \eqref{eq:ident_ymmse} again but this time w.r.t 
$\phi_{\nu i}\!\sim\!{\cal N}(0,1/L)$, one finds 
\begin{align}
 Y_{t,h}^{({\cal S})}=\frac{M^{-u}}{L} \sum_{\nu\in{\cal S}}\sum_{i=1}^N \EE[Z_{{\nu}} \langle U_{\nu} {\bar  X}_i{\bar  X}_i'\rangle_{t,h}-Z_{{\nu}}\langle U_{\nu} {\bar X}^2_i\rangle_{t,h}]\nonumber
\end{align}
(where $\bar \bX=\bX-\bS,\bar \bX'=\bX'-\bS$ and $\bX,\bX'$ 
are i.i.d replicas). Then using \eqref{consequence3} for the first term in the bracket and the definition of $u_\nu$ for the second one, simple algebra leads to $Y_{t,h}^{({\cal S})} \!=\!Y_1 \!-\!Y_2$ where
\begin{align}
Y_1&= \EE[(M^{-u}\sum_{\nu\in{\cal S}}Z_{\nu}^2)  \langle {\cal E}\rangle_{t,h}], \nonumber\\
Y_2&= \sqrt{\frac{t}{\Delta}} M^{-u}\sum_{\nu\in{\cal S}} \EE[Z_{\nu}\langle [\bm{\Phi} \bar \bX]_{\nu} {\cal E}\rangle_{t,h}]. \nonumber
\end{align}
The noise $\bz$ has i.i.d standardized Gaussian components, so the central limit theorem implies
\be
Y_1 = \EE[\langle {\cal E}\rangle_{t,h}](1\!+\!{\cal O}(L^{-u/2}))\!=\!E_{t,h}\!+\!{\cal O}(L^{-u/2}). 
\nonumber
\ee
Below we show that Lemma \ref{concentration} implies for a.e. $h$,
\begin{align}
Y_{2} &=\sqrt{\frac{t}{\Delta}} M^{-u}\sum_{\nu\in{\cal S}} \EE[Z_{\nu}\langle [\bm{\Phi} \bar \bX]_{\nu}\rangle_{t,h}]E_{t,h} \!+\!\smallO_L(1)
\label{eq:Y2afterDecoup}.
\end{align}
Then from \eqref{eq:ident_ymmse} and \eqref{eq:Y2afterDecoup} we get ${Y}_{2} = (t/\Delta) Y_{t,h}^{({\cal S})}E_{t,h}\!+\!\smallO_L(1)$. Putting all pieces together we get 
%
%
\begin{align}
Y_{t,h}^{({\cal S})} = E_{t,h} - (t/\Delta)Y_{t,h}^{({\cal S})} E_{t,h} +\smallO_L(1), \nonumber
\end{align}
which is equivalent to \eqref{Yresult} in Lemma~\ref{lemma:MMSE}.

It remains to justify \eqref{eq:Y2afterDecoup}. We have 
$\EE[Z_{\nu}\langle [\bm{\Phi} \bar \bX]_{\nu} {\cal E}\rangle_{t,h}] = 
\EE[Z_{\nu}\langle [\bm{\Phi} \bar \bX]_{\nu} \rangle_{t,h}]E_{t,h} +
\EE[Z_{\nu}\langle [\bm{\Phi} \bar \bX]_{\nu} \delta{\cal E}\rangle_{t,h}]$. Thus it suffices to show that
the second term is $\smallO_L(1)$ for a.e. $h$. 
From Cauchy-Schwarz
\begin{align}
\EE[Z_{\nu}\langle [\bm{\Phi} \bar \bX]_{\nu} \delta{\cal E}\rangle_{t,h}]^2
\leq 
\EE[\langle \delta{\cal E}^2 \rangle_{t,h}] \EE[Z_{\nu}^2\langle [\bm{\Phi} \bar \bX]_{\nu}^2\rangle_{t,h}] .
\end{align}
As remarked below it, Lemma \ref{concentration} implies $\EE[\langle \delta{\cal E}^2 \rangle_{t,h}] = \smallO_L(1)$
for a.e. $h$, thus we just have to argue that $\EE[Z_{\nu}^2\langle [\bm{\Phi} \bar \bX]_{\nu}^2\rangle_{t,h}]$
is bounded uniformly in $L$. By Cauchy-Schwarz again the square of this quantity is smaller than
\be
 \EE[Z_{\nu}^4] \EE[\langle [\bm{\Phi} \bar \bX]_{\nu}^4\rangle_{t,h}] 
 = 3 \EE[\langle [\bm{\Phi} \bar\bX]_{\nu}^4\rangle_{t,h}]. \label{37}
\ee
Expanding $[\bm{\Phi} \bar\bX]_{\nu}^4$ only terms of the form
$
\mathbb{E}[[\bm{\Phi}\bS]_\nu^n\langle[\bm{\Phi}\bX]_\nu^m \rangle_{t,h}]
$
remain (with $0\leq m,n\le 4$). By Cauchy-Schwarz once more, their square is less than 
\be
\mathbb{E}[[\bm{\Phi}\bS]_\nu^{2n}]\mathbb{E}[\langle[\bm{\Phi}\bX]_\nu^{2m} \rangle_{t,h}] =\!
\mathbb{E}[[\bm{\Phi}\bS]_\nu^{2n}]\mathbb{E}[[\bm{\Phi}\bS]_\nu^{2m}], 
\label{23}
\ee
where the equality comes form the Nishimori identity \eqref{nishibasic}.
It is clear that these moments are all bounded uniformly in $L$. Indeed
$\phi_{\mu i}\!\sim\! \mathcal{N}(0, 1/L)$ is independent of $\bs$, so conditional on $\bs$, the linear combination 
$\bm{\Phi}\bs$ is a Gaussian variable with a variance less than $Bs_{\rm max}^2$. 

The proof of Lemma \ref{lemma:MMSE} is now complete.

\subsection{Proof of Lemma~\ref{lemma:smallEdiff}}
Recall \eqref{consequence2}. Then the MMSE difference can be written as
\begin{align}
|E_{t,h}- &E_{0,h}| =\Big\vert\int_{0}^t\! ds \frac{d}{ds} \EE[\langle{\cal E} \rangle_{s,h} ]\Big\vert
\nonumber\\&
= \Big\vert\sum_{\nu \in\mathcal{S}}\int_{0}^t\! d{s}\, \EE[\langle{\cal E}G_\nu\rangle_{s,h}\!- \!\langle{\cal E}\rangle_{s,h}\langle G_\nu\rangle_{s,h}]\Big\vert ,
\label{19}
\end{align}
where $G_\nu \! \defeq\! ([\bm{\phi}\bar \bx]_\nu^2 - [\bm{\phi}\bar \bx]_\nu z_\nu \sqrt{\Delta/s})/(2\Delta)$.
Note that in \eqref{19} $\mathcal{E}$ can be replaced by $\delta\mathcal{E}$. Also, all 
$G_\nu$'s are statistically equivalent and we can replace them by the first term in the set $\mathcal{S}$, say $\nu=1$. Thus
\begin{align}
&|E_{t,h}\!-\!  E_{0,h}| \leq M^u \!\int_0^t ds\big| \EE[\langle \delta\mathcal{E} G_1\rangle_{s,h}\!-\!\langle\delta\mathcal{E} \rangle_{s,h}\langle G_1\rangle_{s,h}]\big|.\nonumber
\end{align} 
Integrating over $h\!\in\! [a,\epsilon]$, applying Fubini and Cauchy-Schwarz, one gets
\begin{align}
\Big(\int_a^\epsilon dh\, |E_{t,h}\!- \!E_{0,h}|\Big)^2
& \leq 
4M^{2u} \int_{0}^t ds\,\int_a^\epsilon dh\, \EE[\langle \delta {\cal E}^2\rangle_{s,h}]
\nonumber \\ &
\times
\int_{0}^t ds\,
\int_a^\epsilon dh\, \EE[\langle G_1^2\rangle_{s,h}]. \nonumber
\label{25}
\end{align}
Proceeding similarly as in the steps \eqref{37}--\eqref{23} one shows that $\EE[\langle G_1^2\rangle_{s,h}] = \mathcal{O}(1)$ (w.r.t $L$). Lemma~\ref{concentration} allows to conclude
\begin{align}
\int_a^\epsilon dh\, |E_{t,h}\!- \!E_{0,h}| = \mathcal{O}(M^u L^{-1/20}) = \mathcal{O}(L^{u-1/20}),\nonumber
\end{align}
which implies Lemma~\ref{lemma:smallEdiff} by Lebesgue's dominated convergence theorem as the integrand is bounded and $u\!<1/20$.
\section{Conclusion}
Let us end by pointing another application of the sub-extensive interpolation method. In \cite{barbier_ieee_replicaCS} it is used to prove the invariance of the MI under spatial coupling in RLE. 
There, one interpolates between a homogeneous measurement matrix and a spatially coupled one. This is done by iteratively removing sub-extensive blocks of lines in the homogeneous matrix and replacing them by ``spatially coupled'' lines.
%
%
Along this process the MI is monotonously varying which leads to useful inequalities. This partly discrete, 
partly continuous interpolation defines a ``family'' of interpolation methods parametrized by $u$ (the sub-extensive block size parameter). 
Roughly speaking our sub-extensive interpolation method ``interpolates'' between the purely global and continuous method 
of Guerra and Toninelli for dense graphical models~\cite{guerra2005introduction} (at $u\!=\!1$) and the combinatorial approach developed for 
sparse graphs by Gamarnik, Bayati and Tetali in~\cite{bayati2013} (at $u\!=\!0$), where a discrete and local interpolation is done ``one constraint at a time'' 
(here one measurement at a time).
\section*{Acknowledgments}
J.B acknowledges SNSF grant no. 200021-156672.
%
%
\bibliographystyle{IEEEtran}
\bibliography{refs}
\end{document}